\documentclass[aps,prl,reprint,superscriptaddress]{revtex4-1}

\usepackage{graphicx}
\usepackage{amsmath}
\usepackage{color}
\usepackage{subfigure}
\begin{document}


\title{Estimating Hidden Asymptomatics, Herd Immunity Threshold and Lockdown Effects using a COVID-19 Specific Model}


\author{Shaurya Kaushal}
 \affiliation{
 Jawaharlal Nehru Centre for Advanced Scientific Research, Jakkur, Bangalore  560064, India}
 \author{Abhineet Singh Rajput}
 \affiliation{Indian Institute of Science, CV Raman Rd, Bengaluru, Karnataka, India 560012.}
 \author{ Soumyadeep Bhattacharya}
 \affiliation{Sankhya Sutra Labs, Manyata Embassy Business Park, Bengaluru, Karnataka, India 560045.}
 \author{M. Vidyasagar}
 \affiliation{Indian Institute of Technology Hyderabad, Kandi 502285, India}
 \author{Aloke Kumar}
 \affiliation{Indian Institute of Science, CV Raman Rd, Bengaluru, Karnataka, India 560012.}
 \author{Meher K. Prakash}
 \email{meher@jncasr.ac.in}
 \affiliation{
 Jawaharlal Nehru Centre for Advanced Scientific Research, Jakkur, Bangalore  560064, India}
 \affiliation{VNIR Biotechnologies Pvt Ltd, Bangalore Bioinnovation Center, Helix Biotech Park,  Electronic City Phase I, Bangalore 560100.}
 \author{Santosh Ansumali}
 \affiliation{
 Jawaharlal Nehru Centre for Advanced Scientific Research, Jakkur, Bangalore  560064, India}
 \affiliation{Sankhya Sutra Labs, Manyata Embassy Business Park, Bengaluru, Karnataka, India 560045.}

\date{\today}

\begin{abstract}
A quantitative COVID-19 model that incorporates hidden asymptomatic patients is developed, and an analytic solution in parametric form is given. The model incorporates the impact of lockdown and resulting spatial migration of population due to announcement of lockdown. A method is presented for estimating the model parameters from real-world data. It is shown that increase of infections slows down and herd immunity is achieved when symptomatic patients are 4-6\% of the population for the European countries we studied, when the total infected fraction is between 50-56 \%.  Finally, a method for estimating the number of  asymptomatic patients, who have been the key hidden link in the spread of the infections, is  presented.
\end{abstract}

\keywords{COVID-19, lockdown, asymptomatics, SAIR model, exact solution}

\maketitle

COVID-19 infections have breached the five million mark, yet there is neither a vaccine nor a scalable treatment in sight \cite{chan2020familial,enserink2020mathematics}. Furthermore, a distinctive feature of the COVID-19, in contrast to other infectious diseases such as Influenza or SARS, is the presence of a large fraction of ``asymptomatic'' patients, who don't have any obvious symptoms but are still capable of infecting susceptible individuals through contacts. However, identifying individuals spreading infections via  the asymptomatic pathway is not easy unless extensive contact tracing and testing are performed. A major challenge is the uncertainty in the estimation of asymptomatic fraction, with estimates ranging from  $41\%$ to $86\%$ of infected \cite{asymptomatic,nishiura2020estimation}.  And along the symptomatic pathway, 44\% of the infections are spread before the onset of symptoms rendering the quarantining people with symptoms less efficient compared to other infectious diseases.\cite{he2020temporal} These challenges have driven governments to implement non-pharmaceutical interventions (NPIs) such as social distancing and partial or full lockdowns \cite{ferguson2020impact}.  An unsaid, \emph{a posteriori}, rationale for these lockdowns is that they provide efficient isolation mechanism for asymptomatic. However, a dearth of quantitative understanding of the effects of the lockdown has triggered debate around the effectiveness, duration and mode (partial vs. full) of lockdown. Thus, it is even suggested that societies should just move in an unhindered manner, towards the attainment of the ``herd-immunity threshold'' \cite{UK-herd}. This threshold is achieved when a sufficiently large proportion of a population becomes immune, and as a result, the disease spread slows down. For COVID-19, estimating the onset of herd immunity remains elusive, and indeed, ascertaining whether herd immunity exists at all! Moreover, high case fatality rate of $3-10\%$ (vs. $0.05\%$ for seasonal influenza) limits the practicality of herd immunity as an effective policy tool. Thus, models that can provide quantitative estimates of the disease spread and the impact of policy measures are expeditiously required.

Similar to other epidemics/pandemic,  three different kinds of models are used for COVID-19: 1) Statistical extrapolation models which fit the observed patterns of infections to make short-term prediction \cite{covid2020forecasting, prakash2020minimal}, 2) Agent based models for a qualitative illustration of microscopic dynamics of  spreading infections  \cite{singh2020age}, and 3) Compartment models which divide the population into groups based on the current different disease state of the individual and model the interaction among them \cite{robinson2013model,rock2014dynamics,adam2020special,enserink2020mathematics}. Since 1927 plague in Mumbai, compartmental models have been a standard guiding tool for policy decisions \cite{ExactSIR}. The spread of flu-like diseases (influenza, SARS, COVID-19 etc) is often modelled using three or four compartments: Susceptible-Infected-Recovered ($SIR$) or Susceptible-Exposed-Infected-Recovered (SEIR). Some variants, also consider theoretically a simple containment option, of quarantining infected persons with symptoms. However, all these models assume that only contact between the $S$ and the $I$ compartments leads to new infections, with the implicit assumption that contact between the $S$ and $E$ compartments does not lead to any infection. In contrast, an asymptomatic patient with COVID-19 can, and does, infect susceptible individuals through contact. Thus, epidemiological models must consider the distinction between asymptomatic and symptomatic. Moreover, models should distinguish between lockdown and quarantine as these are two qualitatively different policy tools the former operating at the level of a society and the latter the level of a few individuals.

In this letter,  we aim to model all these novel aspects of COVID-19 and to accomplish three goals:
\begin{enumerate}
\item Formulate a minimal epidemiological model incorporating above mentioned  unique aspects of COVID-19 disease spread and associated policies.
\item Establish that the model representatively captures the observed epidemiological data, and sheds light on the underlying parameters and universalities that govern the dynamics in the different phases of the pandemic spread and containment.
\item Use the model to address pertinent questions beyond what is readily measurable -- estimates of the hidden asymptomatics or at what fraction of symptomatic infections herd immunity would be achieved.
\end{enumerate}

We accomplish these objectives by introducing the $SAIR$ model which treats infections by an asymptomatic ($A$) or an infected symptomatic person ($I$) as being equally likely. The dynamical behavior of this model is quite different from that of the $SEIR$ model. The model takes into account lockdown in an explicit fashion by using discontinuous in time reproduction rate (the effective rate at which susceptible population get converted into infected). We give an implicit closed-form solution for this SAIR model, which sheds light on the dynamics of the SAIR model, and also leads to methods for estimating the parameters therein. In order to make this parametric form readily computable, we also introduce an approximate explicit representation. Then we provide a method for estimating the parameters in the model based on the evolution of the disease, and extract the underlying country-specific parameters from the infection data. Further, we show that there exists an intermediate regime immediately after the lockdown that is country-specific, and that the country specific metrics of the success of lockdown can be extracted and analyzed. Then we show that the herd immunity for COVID-19 is achieved when the total symptomatic infections are only around  $5-10\%$ of the population, which is lower than estimated. 

\begin{figure}[h]
	\includegraphics[scale=0.5]{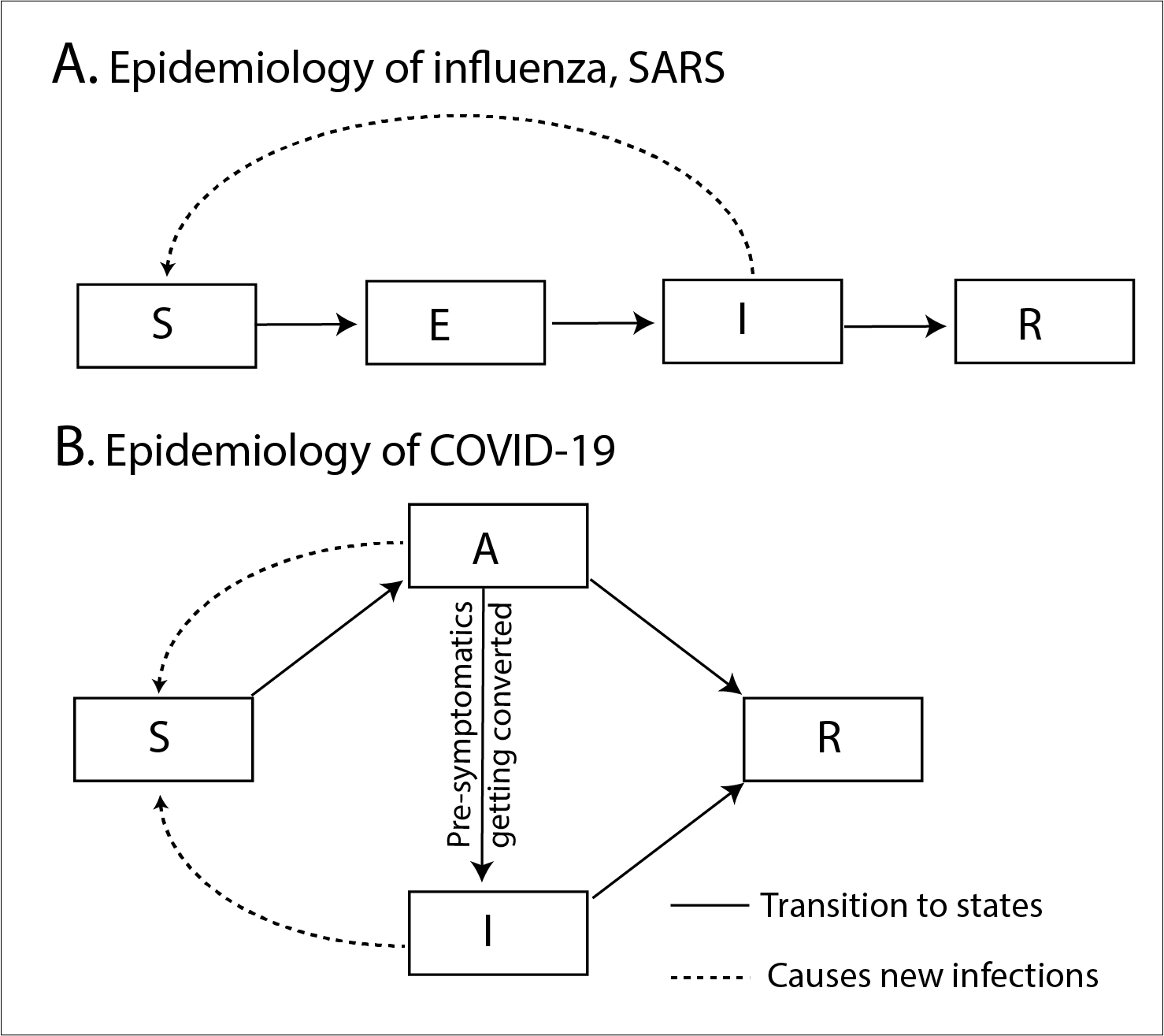}
	\caption{\label{fig1}Schematic of SEIR and SAIR models}
\end{figure}

We begin by emphasizing the difference between $SEIR$ and $SAIR$ models \cite{robinson2013model}.  A typical $SEIR$ model assumes a framework of serial, directed transitions across the intermediate health states of the individuals (FIG.\ref{fig1}). In this framework, the infections are caused when a susceptible person comes in contact with a person deemed to be infected person on the basis of the symptoms (I). However, after this contact, with a certain likelihood the person remains in a pre-symptomatic intermediate state or the exposed individual ($E$), that is not contagious, before transitioning to a contagious and symptomic state ($I$). While this framework is acceptable for influenza or SARS, the epidemiology of COVID-19 is such that there is an alternative pathway between the susceptible ($S$) and the recovered states ($R$) which passes through asymptomatic individuals (estimated to be around 86\%),\cite{asymptomatic} who never show any symptoms but carry enough viral load to infect others. Thus a model for COVID-19 should consider two parallel pathways of infection (Figure 1B).
   
We consider a generalized version of $SAIR$ model as representation of a homogenously mixed population segment  where COVID-19 is spreading. The system will obey the following $SAIR$ dynamics
\begin{align}
\label{Basic1}
 \begin{split}
\dot{S} &= -\alpha(t) \, S  \left(I+A\right),\\
\dot{A} &=  \alpha(t) \,S  \left(  I+   A\right) -\delta\, A-\gamma A+\beta(t) A,
  \\
 \dot{I} &=\delta\, A-\gamma I,\\
 \dot{R} &= \gamma \, (I+A).
 \end{split}
\end{align}
where for any variable $X$ time derivative is denoted as  $\dot{X} =dX/dt$. We assume that  $\alpha(t)$ denotes the probability with which, when a susceptible person meets an infected or asymptomatic person, they become a part of the asymptomatics, which for simplicity includes the pre-symptomatics and the asymptomatics. In our formulation of the model, we claim that the lockdown can be modeled by considering a sudden change in the infection rate constant using a Heavyside function as $\alpha(t)=\alpha_0 {\cal H}(t_{\rm lock} - t )$.  Here, we note in passing that one can model social distancing as reduction in value of $\alpha$ or an imperfect lockdown. In a minimal model, one may assume that asymptomatic patients either get converted into symptomatic one with an effective rate $\delta$ or recovers with a rate $\gamma$. 
This term, typically absent in standard models, denotes the fact that in an idealized lock-down no susceptible  person meets an infected person and thus first order reaction changes to a zero-order reaction.  

Before we proceed to analyse the model, we wish to point out that one may add further complication to this model by introducing more parameters and compartments. For example, recovery rate $\gamma$ and infection rate $\alpha$ need not be same for asymptomatic and symptomatic fraction \cite{robinson2013model}. However, as there is no biological evidence to the contrary, we assume that both rates are equal, which leads to an analytically tractable and simplified framework. 

However, in reality for a large country it is unrealistic to consider it as a homogenously mixed population. Further, during this crisis we learnt that once a lockdown is announced, people migrate across different segments of a country. Even for a qualitatively correct modeling of disease spread dynamics, it is important to account for this migration of people. This migration can indeed happen in many waves. However, for simplicity we assume that it happens once and only during a short duration after lockdown. Furthermore, one would expect that among infected population only asymptomatic people are able to travel.  Here, it needs to be reminded that, we are only interested in the influx of the infected population in a given population segment, and not the details of where they came from. In order to model such a scenario, we take typical thermodynamic route of dividing the system into two parts: system and universe.
Finally, the coupling constant $\beta(t) =  \beta_1 \left\{{\cal H}(t_{\rm lock}+\epsilon - t )-{\cal H}(t_{\rm lock} - t )\right\}$ and $\epsilon$ is the short period of time post lock-down, in which population migration is allowed/possible. This migration is a characteristic of the system (country or region under consideration) and parameters $\beta$ and $\epsilon$ need to be extracted from the data. The rest of the world can also be assumed for this purpose to be following a similar $SAIR$ dynamics
\begin{align}
\label{Basic2}
 \begin{split}
   \dot{S}_{\rm U} &= -\alpha(t) S_{\rm U}  \, \left(I_{\rm U}+A_{\rm U}\right) ,\\
  \dot{A}_{\rm U} &=  \alpha(t) S_{\rm U} \, \left(  I_{\rm U}+   A_{\rm U}\right) -\delta\,  A_{\rm U}-\gamma A_{\rm U}-\beta A,
  \\
  \dot{I}_{\rm U} &=\delta  A_{\rm U}-\gamma I_{\rm U},\\
  \dot{R}_{\rm U} &= \gamma \, (I_{\rm U}+A_{\rm U}).
 \end{split}
\end{align}
 Eq.\eqref{Basic1} and \eqref{Basic2} complete our development of COVID-19 specific model. In the present work, we solved a phenomenological model of a well-mixed society, with everyone interacting with everyone else. However, the interactions may be structured by age, local movement of the population, and many of these can be modelled in the framework of agent based models. The formulation of the disease specific interactions we developed can also be integrated into other models which study the interactions at agent level detail, or in tandem with economic consequences \cite{li2019affine}, both of which are beyond the scope of the present work. With an emphasis mainly on the spread of infections at the societal level, we show that the set of equations we model are sufficient to capture most of the available epidemiological data on COVID-19.

	This system of equations can be solved for pre-lockdown  situation  in terms
	of the reproduction rate $r_0=\alpha_0/\gamma$ by defining $M = I+A$,
	and observing that before lockdown,  we have
	\begin{equation}
	\frac{d \log{S}}{dR} = -r_0, \qquad 
	\frac{d M}{d {S}}= -1+\frac{ 1 }{S r_0} .
	\end{equation}
	which  can be solved  in terms of $\tilde{S}=S/S_0$ as
	\begin{equation}
	\label{First_step}
	R = - r_0^{-1} \log{\tilde{S}}, \qquad \quad
	M=1-S+r_0^{-1}\log{\tilde{S}}
	\end{equation}
	where $M+S+R=1$ at any instant, $S_0$ denotes the susceptible population at
	$t=0$ and the recovered population at $t=0$ is taken to be $0$.
	\\
	On the other hand, after an idealized lock-down no susceptible
	person meets an infected person and thus the first order reaction changes to
	a zero order reaction.
	The intermediate time ($t_{\rm lock}<t<t_{\rm lock} + \epsilon$) solution  simplifies to
	\begin{equation}
	M = \exp(-\gamma t ) \left[  I_{\rm lock} + A_{ \rm lock} (1 + \delta) \exp ((\beta_1-\delta)t )  \right]
	\end{equation}
	Once there is no more flux of asymptomatic individuals, the equations for $M$ yield an exponential decay given by
	\begin{align}
	\label{recovery_rate_post_lock}
	M = M_{\rm lock + \epsilon}\> \> \exp(-\gamma t), \qquad \quad 	
	R = 1 - M - S_{\rm lock}
	\end{align}
	Substituting the expression from Eq.\eqref{First_step}  in the evolution equation for $S$ 
	gives us the parametric solution in implicit form as
	\begin{equation}
	\label{implicit}
	\alpha_0\, t= \int_{1}^{\tilde{S}}\frac{d s}{ s  \, \left(-1+S_0 s- r_0^{-1}
		\log{s}\right) } 
	\end{equation}
	Assuming that the equation can be converted to an explicit form for
	$\tilde{S}$ as a function of $t$, it is possible to substitute this into
	Eq.\eqref{First_step} to obtain an expression for $M$ as a function of $t$.
	Finally, the expression for $M(t)$ can be disambiguated into separate
	expressions for $I(t)$ and $A(t)$ by using Eq.\eqref{Basic1}.
	Specifically, in the equation for $\dot{I}$, we can substitute $A = M-I$,
	which gives
	\begin{displaymath}
	\dot{I} = - \left( \delta + \gamma \right) I(t)
	+ \delta M(t) .
	\end{displaymath}
	If we define a new constant $\delta_1 = \gamma + \delta $, then the
	solution of the above equation is
	\begin{equation}
	I(t) = \exp(-\delta_1 t) \left[ I_0 + \delta \int_0^t M(s) \exp(\delta_1 s) ds \right]
	\end{equation}
Therefore the key is to turn Eq.\eqref{implicit} into an explicit expression,
	to the extent possible.
	For this purpose,
	we use Hermite-Hadamard inequality for the logarithm \cite{atif}
	\begin{equation}
	\frac{z-1}{z}\leq \log{z}\leq z-1
	\end{equation}
	which suggests that we use approximate form of the logarithm as $\log{z}=(z-1)(w_1/z +  w_2)$,  with the constraint that $w_1 + w_2 = 1$. Upon approximating the logarithm, we  get a solution in explicit form as 
	\begin{equation}
	\tilde{S} = \frac{h   (1+ h_2 \exp(h \, \alpha_0 t)) }{2a   \left(1-h_2 \exp(h \, \alpha_0 t)\right)}
	+\frac{b}{2a}
	\end{equation}
	where $ a = (S_0 r_0 -  w_2)/ r_0, \; 
	b=   \left(r_0 +  w_1-  w_2\right)/r_0, \,  d=  w_1/{r_0}$.
	where $h_2 =    (2\, a-b-h)/(2\, a-b+h)$,  and $h$ is a constant such that and $h=\sqrt{b^2-4ad}$. Once the evolution equation for $S$ is known in a closed from, we find the evolution for the remaining variables using Eq. \eqref{First_step}. FIG \ref{fig2} depicts a representative temporal variation for the parameters $S$,$A$, $I$ and $R$ captured using the analytical solution. The analytical solution formulated using the above approximation to logarithm is found to be in close agreement with the numerical solution of the ODE (see Supplemental Material \cite{supp}).     
\begin{figure}[h]
	\subfigure[]{
	\includegraphics[scale=0.15]{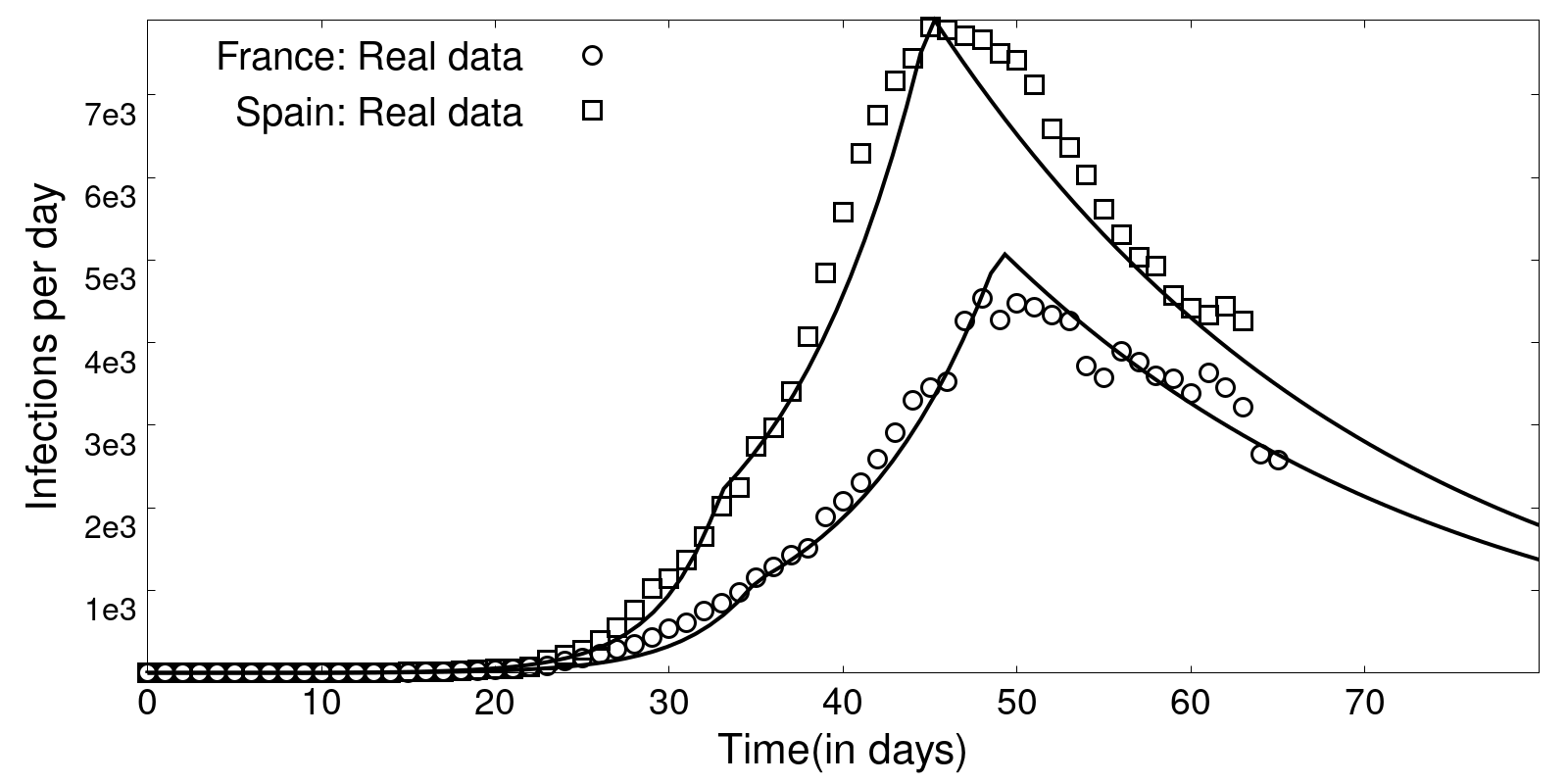}}
    \subfigure[]{
	\includegraphics[scale=0.15]{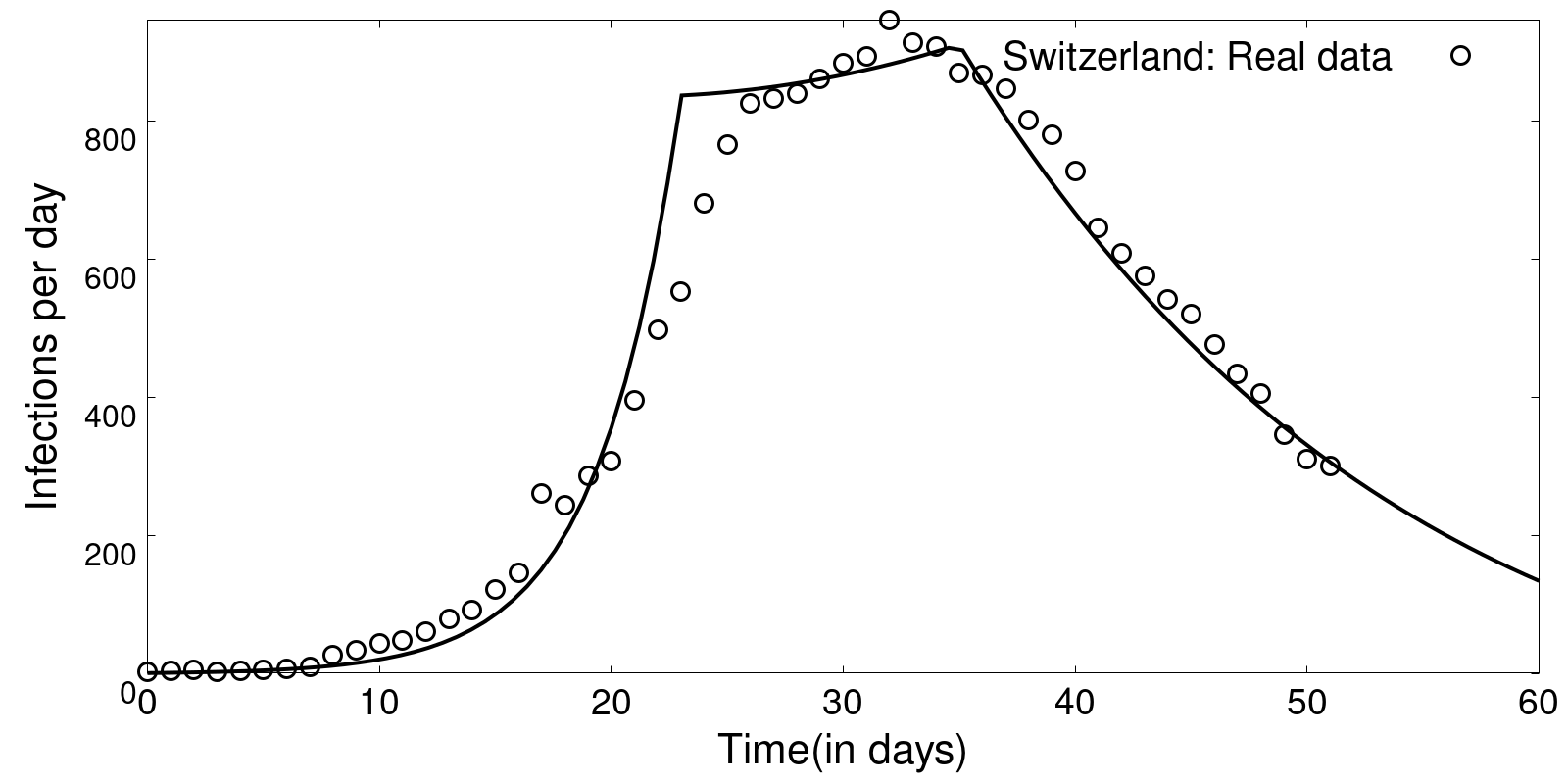}}
\caption{\label{fig3} Fits to the infection data from (a) France and Spain, (b) Switzerland Since the data is stochastic, a 3-day average was used for obtaining the fits. The extracted parameters are tabulated in Supplementary Table 1 } 
\end{figure}

The evolution of infections pre-lockdown and in early time limit is given by
\begin{align}
\label{pre_lock_infection}
I = \exp \{ -\delta_1 t \} \left[I_0 +  \int_0^t ds  \left( \frac{ \delta \exp\{ \delta_1 s \}}{ r_0} \right)
\left( k - g\tilde{S}(s) \right) \right]
\end{align}
where $k=(r_0-1)$ and $g=(S_0r_0-1)$. The solution post lock-down is given by
\begin{align}
\label{infections_post_lock}
\begin{split}
I &= \exp \{-\gamma t\}(I_{\rm lock} - L)
\\
& +   L \exp\{ (-\delta_1 + \beta_1(1-{H}(t-\epsilon)))t \}
\end{split}
\end{align}
where, 
\begin{align}
 L =  \frac{\delta \left( A_{\rm lock} \> \exp \{ \epsilon(\beta_1 - \delta_1) {H}(t-\epsilon) \} \right) } {(\gamma-\delta_1)+\beta_1(1-{H}(t-\epsilon))}
\end{align}

Eqs. \eqref{pre_lock_infection}, Eq.\eqref{recovery_rate_post_lock} and Eq.\eqref{infections_post_lock} are the closed-form solutions to the model we developed. Epidemics like SARS in 2003, Swine flu in 2009, MERS in 2012 and 2015, could be managed at most with contact tracing and quarantine, and hence addressing a solution for the lockdown did not arise. COVID-19 thus presented itself with the unique infection scenarios and the challenges of the lockdown for its mitigation, and our model and its closed form-solution address these uniquene aspects.

The reported infection data from different countries had three regimes - rising, intermediate and decreasing, if they implemented a lockdown. It can be easily assumed that the reported infections are the symptomatic infections, since most countries have been short of testing resources; as a result, patients were tested for a confirmation only after the onset of symptoms. The equations derived above for $I$ could be fit will all these three different regimes. In the process, we could extract the governing parameters. The parameter $\alpha_0$, $\gamma$ and $\delta_1$ are estimated by fitting Eq. \eqref{pre_lock_infection}, Eq.\eqref{recovery_rate_post_lock} and Eq.\eqref{infections_post_lock} respectively, to the publicly available data pertaining to the pre and post lock-down period for various countries (see Supplemental Material \cite{supp}). The parameters $\alpha_0$ representing the rise is similar for many countries reiterating a universal pattern in the initial pre-lockdown regime. This can be understood as an intrinsic characteristic dynamics of COVID-19 which exhibits strong similarities across countries (see Table 1 in Supplemental Material \cite{supp}). But a much stronger country-specific disease dynamics was the intermediate regime, described using the parameter $\beta_1$. The formal solution in (Eq. \ref{infections_post_lock}) is fit to the infection rate right after lock-down to estimate the parameter $\beta_1$ and $\alpha$ (see Supplementary FIGS.3,4). This is to be expected as migration during lockdown can be expected to be a country-specific event dictated by the prevalent social-political conditions. 
 

\begin{figure}[h]
	\includegraphics[scale=0.15]{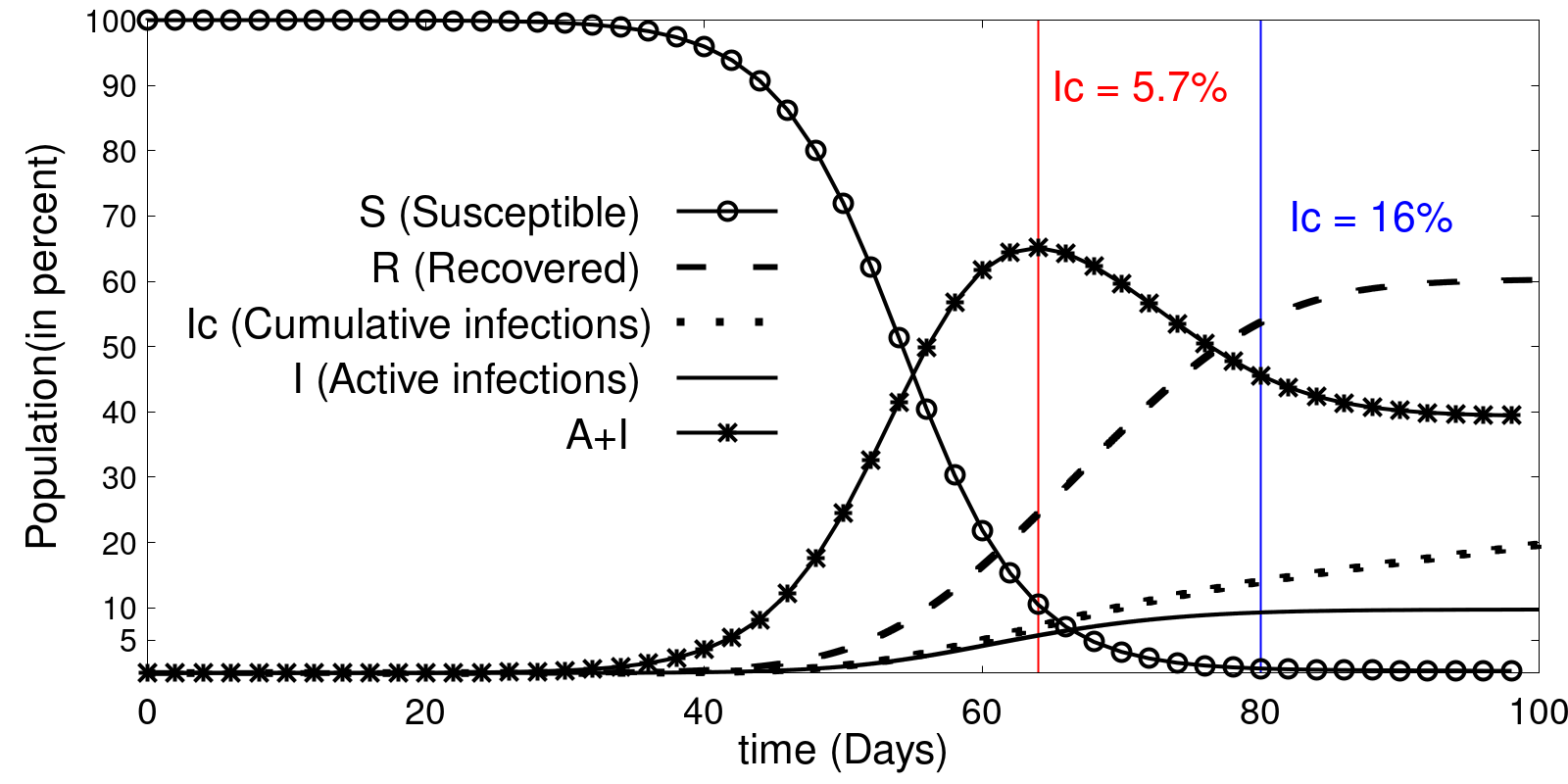}
	\caption{\label{fig2} Analytical solution of the SAIR model using parameters $\alpha=0.25, \gamma = 0.027, \delta_1 = 0.036$, which are in reasonable range of real time parameter values for COVID19 (see Supplemental Material). The blue and the red lines indicate the maxima, considering only the symptomatic or the total infections respectively. The infection rate slows down significantly and a herd-immunity is achieved after the combined infections reach a peak when the symptomatic infections have crossed $\approx 5.7\%$ of the total population. The peak of the symptomatic infections is achieved at around 16\%, just before the final saturation value of around 20\%.}
\end{figure}

  

With these validations for the levels of infections $I$ that were observed in the different countries, and the parameters that were extracted, we could estimate how the number of individuals in the individual compartments $S$, $A$, $I$ and $R$ changed with time with or without a lockdown (FIG.\ref{fig2}). Because it had been impossible to test the entire population or even a significant fraction of it, the asymptomatics have remained a missing link in the epidemiology, although certain estimates suggest a 1:10 ratio between the sympomatic and asymptomatic individuals.  Using our model, we could estimate the ratio of the asymptomatic to symptomatic individuals (Supplementary Fig. 5), which varies from 1 to 40 depending on the phase of the pandemic dynamics. Our results show that the herd-immunity, defined as the fraction of population at which symptomatic infections reach a peak and beyond which begin decreasing could be achieved at 4-6\% of the population as illustrated in FIG.\ref{fig2}) (TABLE 2 in Supplementary Information).These estimates for herd-immunity which are in single digit percentages only seem contradictory to estimates of 50-60\% \cite{randolph2020herd} until one realises the large fraction of the infections are asymptomatic accounting for a total infection of 50-56\% of the population (Supplementary Table 2). Thus our model allowed us to make estimates both for the hidden-asymptomatics and the herd-immunity, and the fraction of the symptomatics who will burden the health care system.

In conclusion, as a part of our analysis, we are able to provide a method for estimating the asymptomatic fraction of the population. Finally, by fitting our model to data from countries where the pandemic appears to have peaked, we are also able to estimate the level of herd-immunity. We are able to show that herd-immunity is achieved at levels of just 5\% to 10\%, far lower than the levels suggested in the literature. We find that the $SAIR$ model can be readily adapted to incorporate the effects of lockdown and the solution to the system of equations bears striking resemblance to the real-world data. The formal solution allows one to evaluate the effect of lockdown as a policy tool and can also be integrated into other frameworks which study the economic consequences of the lockdowns.

{\bf Acknowledgements.} SA and MKP would like to thank Prof. Srikanth Sastry for helpful discussions. MV would like to thank SERB for funding.

\bibliography{MS}

\begin{thebibliography}{18}%
\makeatletter
\providecommand \@ifxundefined [1]{%
 \@ifx{#1\undefined}
}%
\providecommand \@ifnum [1]{%
 \ifnum #1\expandafter \@firstoftwo
 \else \expandafter \@secondoftwo
 \fi
}%
\providecommand \@ifx [1]{%
 \ifx #1\expandafter \@firstoftwo
 \else \expandafter \@secondoftwo
 \fi
}%
\providecommand \natexlab [1]{#1}%
\providecommand \enquote  [1]{``#1''}%
\providecommand \bibnamefont  [1]{#1}%
\providecommand \bibfnamefont [1]{#1}%
\providecommand \citenamefont [1]{#1}%
\providecommand \href@noop [0]{\@secondoftwo}%
\providecommand \href [0]{\begingroup \@sanitize@url \@href}%
\providecommand \@href[1]{\@@startlink{#1}\@@href}%
\providecommand \@@href[1]{\endgroup#1\@@endlink}%
\providecommand \@sanitize@url [0]{\catcode `\\12\catcode `\$12\catcode
  `\&12\catcode `\#12\catcode `\^12\catcode `\_12\catcode `\%12\relax}%
\providecommand \@@startlink[1]{}%
\providecommand \@@endlink[0]{}%
\providecommand \url  [0]{\begingroup\@sanitize@url \@url }%
\providecommand \@url [1]{\endgroup\@href {#1}{\urlprefix }}%
\providecommand \urlprefix  [0]{URL }%
\providecommand \Eprint [0]{\href }%
\providecommand \doibase [0]{https://doi.org/}%
\providecommand \selectlanguage [0]{\@gobble}%
\providecommand \bibinfo  [0]{\@secondoftwo}%
\providecommand \bibfield  [0]{\@secondoftwo}%
\providecommand \translation [1]{[#1]}%
\providecommand \BibitemOpen [0]{}%
\providecommand \bibitemStop [0]{}%
\providecommand \bibitemNoStop [0]{.\EOS\space}%
\providecommand \EOS [0]{\spacefactor3000\relax}%
\providecommand \BibitemShut  [1]{\csname bibitem#1\endcsname}%
\let\auto@bib@innerbib\@empty
\bibitem [{\citenamefont {Chan}\ \emph {et~al.}(2020)\citenamefont {Chan},
  \citenamefont {Yuan}, \citenamefont {Kok}, \citenamefont {To}, \citenamefont
  {Chu}, \citenamefont {Yang}, \citenamefont {Xing}, \citenamefont {Liu},
  \citenamefont {Yip}, \citenamefont {Poon} \emph {et~al.}}]{chan2020familial}%
  \BibitemOpen
  \bibfield  {author} {\bibinfo {author} {\bibfnamefont {J.~F.-W.}\
  \bibnamefont {Chan}}, \bibinfo {author} {\bibfnamefont {S.}~\bibnamefont
  {Yuan}}, \bibinfo {author} {\bibfnamefont {K.-H.}\ \bibnamefont {Kok}},
  \bibinfo {author} {\bibfnamefont {K.~K.-W.}\ \bibnamefont {To}}, \bibinfo
  {author} {\bibfnamefont {H.}~\bibnamefont {Chu}}, \bibinfo {author}
  {\bibfnamefont {J.}~\bibnamefont {Yang}}, \bibinfo {author} {\bibfnamefont
  {F.}~\bibnamefont {Xing}}, \bibinfo {author} {\bibfnamefont {J.}~\bibnamefont
  {Liu}}, \bibinfo {author} {\bibfnamefont {C.~C.-Y.}\ \bibnamefont {Yip}},
  \bibinfo {author} {\bibfnamefont {R.~W.-S.}\ \bibnamefont {Poon}}, \emph
  {et~al.},\ }\href@noop {} {\bibfield  {journal} {\bibinfo  {journal} {The
  Lancet}\ }\textbf {\bibinfo {volume} {395}},\ \bibinfo {pages} {514}
  (\bibinfo {year} {2020})}\BibitemShut {NoStop}%
\bibitem [{\citenamefont {Enserink}\ and\ \citenamefont
  {Kupferschmidt}(2020)}]{enserink2020mathematics}%
  \BibitemOpen
  \bibfield  {author} {\bibinfo {author} {\bibfnamefont {M.}~\bibnamefont
  {Enserink}}\ and\ \bibinfo {author} {\bibfnamefont {K.}~\bibnamefont
  {Kupferschmidt}},\ }\href@noop {} {\bibfield  {journal} {\bibinfo  {journal}
  {Science Magazine}\ } (\bibinfo {year} {2020})}\BibitemShut {NoStop}%
\bibitem [{\citenamefont {Li}\ \emph {et~al.}(2020)\citenamefont {Li},
  \citenamefont {Pei}, \citenamefont {Chen}, \citenamefont {Song},
  \citenamefont {Zhang}, \citenamefont {Yang},\ and\ \citenamefont
  {Shaman}}]{asymptomatic}%
  \BibitemOpen
  \bibfield  {author} {\bibinfo {author} {\bibfnamefont {R.}~\bibnamefont
  {Li}}, \bibinfo {author} {\bibfnamefont {S.}~\bibnamefont {Pei}}, \bibinfo
  {author} {\bibfnamefont {B.}~\bibnamefont {Chen}}, \bibinfo {author}
  {\bibfnamefont {Y.}~\bibnamefont {Song}}, \bibinfo {author} {\bibfnamefont
  {T.}~\bibnamefont {Zhang}}, \bibinfo {author} {\bibfnamefont
  {W.}~\bibnamefont {Yang}},\ and\ \bibinfo {author} {\bibfnamefont
  {J.}~\bibnamefont {Shaman}},\ }\bibfield  {journal} {\bibinfo  {journal}
  {Science}\ }\href {https://doi.org/10.1126/science.abb3221}
  {10.1126/science.abb3221} (\bibinfo {year} {2020})\BibitemShut {NoStop}%
\bibitem [{\citenamefont {Nishiura}\ \emph {et~al.}(2020)\citenamefont
  {Nishiura}, \citenamefont {Kobayashi}, \citenamefont {Miyama}, \citenamefont
  {Suzuki}, \citenamefont {Jung}, \citenamefont {Hayashi}, \citenamefont
  {Kinoshita}, \citenamefont {Yang}, \citenamefont {Yuan}, \citenamefont
  {Akhmetzhanov} \emph {et~al.}}]{nishiura2020estimation}%
  \BibitemOpen
  \bibfield  {author} {\bibinfo {author} {\bibfnamefont {H.}~\bibnamefont
  {Nishiura}}, \bibinfo {author} {\bibfnamefont {T.}~\bibnamefont {Kobayashi}},
  \bibinfo {author} {\bibfnamefont {T.}~\bibnamefont {Miyama}}, \bibinfo
  {author} {\bibfnamefont {A.}~\bibnamefont {Suzuki}}, \bibinfo {author}
  {\bibfnamefont {S.}~\bibnamefont {Jung}}, \bibinfo {author} {\bibfnamefont
  {K.}~\bibnamefont {Hayashi}}, \bibinfo {author} {\bibfnamefont
  {R.}~\bibnamefont {Kinoshita}}, \bibinfo {author} {\bibfnamefont
  {Y.}~\bibnamefont {Yang}}, \bibinfo {author} {\bibfnamefont {B.}~\bibnamefont
  {Yuan}}, \bibinfo {author} {\bibfnamefont {A.~R.}\ \bibnamefont
  {Akhmetzhanov}}, \emph {et~al.},\ }\href@noop {} {\bibfield  {journal}
  {\bibinfo  {journal} {medRxiv}\ } (\bibinfo {year} {2020})}\BibitemShut
  {NoStop}%
\bibitem [{\citenamefont {He}\ \emph {et~al.}(2020)\citenamefont {He},
  \citenamefont {Lau}, \citenamefont {Wu}, \citenamefont {Deng}, \citenamefont
  {Wang}, \citenamefont {Hao}, \citenamefont {Lau}, \citenamefont {Wong},
  \citenamefont {Guan}, \citenamefont {Tan} \emph {et~al.}}]{he2020temporal}%
  \BibitemOpen
  \bibfield  {author} {\bibinfo {author} {\bibfnamefont {X.}~\bibnamefont
  {He}}, \bibinfo {author} {\bibfnamefont {E.~H.}\ \bibnamefont {Lau}},
  \bibinfo {author} {\bibfnamefont {P.}~\bibnamefont {Wu}}, \bibinfo {author}
  {\bibfnamefont {X.}~\bibnamefont {Deng}}, \bibinfo {author} {\bibfnamefont
  {J.}~\bibnamefont {Wang}}, \bibinfo {author} {\bibfnamefont {X.}~\bibnamefont
  {Hao}}, \bibinfo {author} {\bibfnamefont {Y.~C.}\ \bibnamefont {Lau}},
  \bibinfo {author} {\bibfnamefont {J.~Y.}\ \bibnamefont {Wong}}, \bibinfo
  {author} {\bibfnamefont {Y.}~\bibnamefont {Guan}}, \bibinfo {author}
  {\bibfnamefont {X.}~\bibnamefont {Tan}}, \emph {et~al.},\ }\href@noop {}
  {\bibfield  {journal} {\bibinfo  {journal} {Nature medicine}\ ,\ \bibinfo
  {pages} {1}} (\bibinfo {year} {2020})}\BibitemShut {NoStop}%
\bibitem [{\citenamefont {Ferguson}\ \emph {et~al.}(2020)\citenamefont
  {Ferguson}, \citenamefont {Walker}, \citenamefont {Whittaker} \emph
  {et~al.}}]{ferguson2020impact}%
  \BibitemOpen
  \bibfield  {author} {\bibinfo {author} {\bibfnamefont {N.}~\bibnamefont
  {Ferguson}}, \bibinfo {author} {\bibfnamefont {P.}~\bibnamefont {Walker}},
  \bibinfo {author} {\bibfnamefont {C.}~\bibnamefont {Whittaker}}, \emph
  {et~al.},\ }\href@noop {} {\emph {\bibinfo {title} {Impact of
  non-pharmaceutical interventions (NPIs) to reduce COVID19 mortality and
  healthcare demand. Imperial College London COVID-19 Reports}}},\ \bibinfo
  {type} {Tech. Rep.}\ (\bibinfo  {institution} {Report},\ \bibinfo {year}
  {2020})\BibitemShut {NoStop}%
\bibitem [{\citenamefont {Vallance}()}]{UK-herd}%
  \BibitemOpen
  \bibfield  {author} {\bibinfo {author} {\bibfnamefont {S.~P.}\ \bibnamefont
  {Vallance}},\ }\href@noop {} {\bibinfo  {journal}
  {https://www.theguardian.com/world/coronavirus science chief defends uk
  measures criticism herd immunity}\ }\BibitemShut {NoStop}%
\bibitem [{\citenamefont {COVID}\ \emph {et~al.}(2020)\citenamefont {COVID},
  \citenamefont {Murray} \emph {et~al.}}]{covid2020forecasting}%
  \BibitemOpen
\bibfield  {journal} {  }\bibfield  {author} {\bibinfo {author} {\bibfnamefont
  {I.}~\bibnamefont {COVID}}, \bibinfo {author} {\bibfnamefont {C.~J.}\
  \bibnamefont {Murray}}, \emph {et~al.},\ }\href@noop {} {\bibfield  {journal}
  {\bibinfo  {journal} {MedRxiv}\ } (\bibinfo {year} {2020})}\BibitemShut
  {NoStop}%
\bibitem [{\citenamefont {Prakash}\ \emph {et~al.}(2020)\citenamefont
  {Prakash}, \citenamefont {Kaushal}, \citenamefont {Bhattacharya},
  \citenamefont {Chandran}, \citenamefont {Kumar},\ and\ \citenamefont
  {Ansumali}}]{prakash2020minimal}%
  \BibitemOpen
  \bibfield  {author} {\bibinfo {author} {\bibfnamefont {M.~K.}\ \bibnamefont
  {Prakash}}, \bibinfo {author} {\bibfnamefont {S.}~\bibnamefont {Kaushal}},
  \bibinfo {author} {\bibfnamefont {S.}~\bibnamefont {Bhattacharya}}, \bibinfo
  {author} {\bibfnamefont {A.}~\bibnamefont {Chandran}}, \bibinfo {author}
  {\bibfnamefont {A.}~\bibnamefont {Kumar}},\ and\ \bibinfo {author}
  {\bibfnamefont {S.}~\bibnamefont {Ansumali}},\ }\href@noop {} {\bibfield
  {journal} {\bibinfo  {journal} {medRxiv}\ } (\bibinfo {year}
  {2020})}\BibitemShut {NoStop}%
\bibitem [{\citenamefont {Singh}\ and\ \citenamefont
  {Adhikari}(2020)}]{singh2020age}%
  \BibitemOpen
  \bibfield  {author} {\bibinfo {author} {\bibfnamefont {R.}~\bibnamefont
  {Singh}}\ and\ \bibinfo {author} {\bibfnamefont {R.}~\bibnamefont
  {Adhikari}},\ }\href@noop {} {\bibfield  {journal} {\bibinfo  {journal}
  {arXiv preprint arXiv:2003.12055}\ } (\bibinfo {year} {2020})}\BibitemShut
  {NoStop}%
\bibitem [{\citenamefont {Robinson}\ and\ \citenamefont
  {Stilianakis}(2013)}]{robinson2013model}%
  \BibitemOpen
  \bibfield  {author} {\bibinfo {author} {\bibfnamefont {M.}~\bibnamefont
  {Robinson}}\ and\ \bibinfo {author} {\bibfnamefont {N.~I.}\ \bibnamefont
  {Stilianakis}},\ }\href@noop {} {\bibfield  {journal} {\bibinfo  {journal}
  {Mathematical biosciences}\ }\textbf {\bibinfo {volume} {243}},\ \bibinfo
  {pages} {163} (\bibinfo {year} {2013})}\BibitemShut {NoStop}%
\bibitem [{\citenamefont {Rock}\ \emph {et~al.}(2014)\citenamefont {Rock},
  \citenamefont {Brand}, \citenamefont {Moir},\ and\ \citenamefont
  {Keeling}}]{rock2014dynamics}%
  \BibitemOpen
  \bibfield  {author} {\bibinfo {author} {\bibfnamefont {K.}~\bibnamefont
  {Rock}}, \bibinfo {author} {\bibfnamefont {S.}~\bibnamefont {Brand}},
  \bibinfo {author} {\bibfnamefont {J.}~\bibnamefont {Moir}},\ and\ \bibinfo
  {author} {\bibfnamefont {M.~J.}\ \bibnamefont {Keeling}},\ }\href@noop {}
  {\bibfield  {journal} {\bibinfo  {journal} {Reports on Progress in Physics}\
  }\textbf {\bibinfo {volume} {77}},\ \bibinfo {pages} {026602} (\bibinfo
  {year} {2014})}\BibitemShut {NoStop}%
\bibitem [{\citenamefont {Adam}(2020)}]{adam2020special}%
  \BibitemOpen
  \bibfield  {author} {\bibinfo {author} {\bibfnamefont {D.}~\bibnamefont
  {Adam}},\ }\href@noop {} {\bibfield  {journal} {\bibinfo  {journal} {Nature}\
  }\textbf {\bibinfo {volume} {580}},\ \bibinfo {pages} {316} (\bibinfo {year}
  {2020})}\BibitemShut {NoStop}%
\bibitem [{\citenamefont {Kermack}\ and\ \citenamefont
  {McKendrick}(1991)}]{ExactSIR}%
  \BibitemOpen
  \bibfield  {author} {\bibinfo {author} {\bibfnamefont {W.~O.}\ \bibnamefont
  {Kermack}}\ and\ \bibinfo {author} {\bibfnamefont {A.~G.}\ \bibnamefont
  {McKendrick}},\ }\href@noop {} {\bibinfo {title} {Contributions to the
  mathematical theory of epidemics--i. 1927.}} (\bibinfo {year}
  {1991})\BibitemShut {NoStop}%
\bibitem [{\citenamefont {Li}\ \emph {et~al.}(2019)\citenamefont {Li},
  \citenamefont {Boghosian},\ and\ \citenamefont {Li}}]{li2019affine}%
  \BibitemOpen
  \bibfield  {author} {\bibinfo {author} {\bibfnamefont {J.}~\bibnamefont
  {Li}}, \bibinfo {author} {\bibfnamefont {B.~M.}\ \bibnamefont {Boghosian}},\
  and\ \bibinfo {author} {\bibfnamefont {C.}~\bibnamefont {Li}},\ }\href@noop
  {} {\bibfield  {journal} {\bibinfo  {journal} {Physica A: Statistical
  Mechanics and its Applications}\ }\textbf {\bibinfo {volume} {516}},\
  \bibinfo {pages} {423} (\bibinfo {year} {2019})}\BibitemShut {NoStop}%
\bibitem [{\citenamefont {Atif}\ \emph {et~al.}(2017)\citenamefont {Atif},
  \citenamefont {Kolluru}, \citenamefont {Thantanapally},\ and\ \citenamefont
  {Ansumali}}]{atif}%
  \BibitemOpen
  \bibfield  {author} {\bibinfo {author} {\bibfnamefont {M.}~\bibnamefont
  {Atif}}, \bibinfo {author} {\bibfnamefont {P.~K.}\ \bibnamefont {Kolluru}},
  \bibinfo {author} {\bibfnamefont {C.}~\bibnamefont {Thantanapally}},\ and\
  \bibinfo {author} {\bibfnamefont {S.}~\bibnamefont {Ansumali}},\ }\href@noop
  {} {\bibfield  {journal} {\bibinfo  {journal} {Physical review letters}\
  }\textbf {\bibinfo {volume} {119}},\ \bibinfo {pages} {240602} (\bibinfo
  {year} {2017})}\BibitemShut {NoStop}%
\bibitem [{sup()}]{supp}%
  \BibitemOpen
  \href@noop {} {\bibinfo  {journal} {See Supplemental Material at}\
  }\BibitemShut {NoStop}%
\bibitem [{\citenamefont {Randolph}\ and\ \citenamefont
  {Barreiro}(2020)}]{randolph2020herd}%
  \BibitemOpen
\bibfield  {journal} {  }\bibfield  {author} {\bibinfo {author} {\bibfnamefont
  {H.~E.}\ \bibnamefont {Randolph}}\ and\ \bibinfo {author} {\bibfnamefont
  {L.~B.}\ \bibnamefont {Barreiro}},\ }\href@noop {} {\bibfield  {journal}
  {\bibinfo  {journal} {Immunity}\ }\textbf {\bibinfo {volume} {52}},\ \bibinfo
  {pages} {737} (\bibinfo {year} {2020})}\BibitemShut {NoStop}%
\end{thebibliography}%

\end{document}


\title{Supplemental Material \\ Estimating Hidden Asymptomatics, Herd Immunity Threshold and Lockdown Effects using a COVID-19 Specific Model}
\author{Shaurya Kaushal}
\affiliation{
	Jawaharlal Nehru Centre for Advanced Scientific Research, Jakkur, Bangalore  560064, India}

\author{Abhineet Singh Rajput}
\affiliation{Indian Institute of Science, CV Raman Rd, Bengaluru, Karnataka, India 560012.}
\author{ Soumyadeep Bhattacharya}
\affiliation{Sankhya Sutra Labs, Manyata Embassy Business Park, Bengaluru, Karnataka, India 560045.}
\author{M. Vidyasagar}
\affiliation{Indian Institute of Technology Hyderabad, Kandi 502285, India}
\author{Aloke Kumar}
\affiliation{Indian Institute of Science, CV Raman Rd, Bengaluru, Karnataka, India 560012.}
\author{Meher K. Prakash}
\email{meher@jncasr.ac.in}
\affiliation{
	Jawaharlal Nehru Centre for Advanced Scientific Research, Jakkur, Bangalore  560064, India}
\affiliation{VNIR Biotechnologies Pvt Ltd, Bangalore Bioinnovation Center, Helix Biotech Park,  Electronic City Phase I, Bangalore 560100.}
\author{Santosh Ansumali}
\affiliation{
	Jawaharlal Nehru Centre for Advanced Scientific Research, Jakkur, Bangalore  560064, India}
\affiliation{Sankhya Sutra Labs, Manyata Embassy Business Park, Bengaluru, Karnataka, India 560045.}

\date{\today}
\maketitle

\section{I. ANALYTICAL SOLUTION, APPROXIMATING LOGARITHM}
\begin{figure}
	\includegraphics[width=8cm]{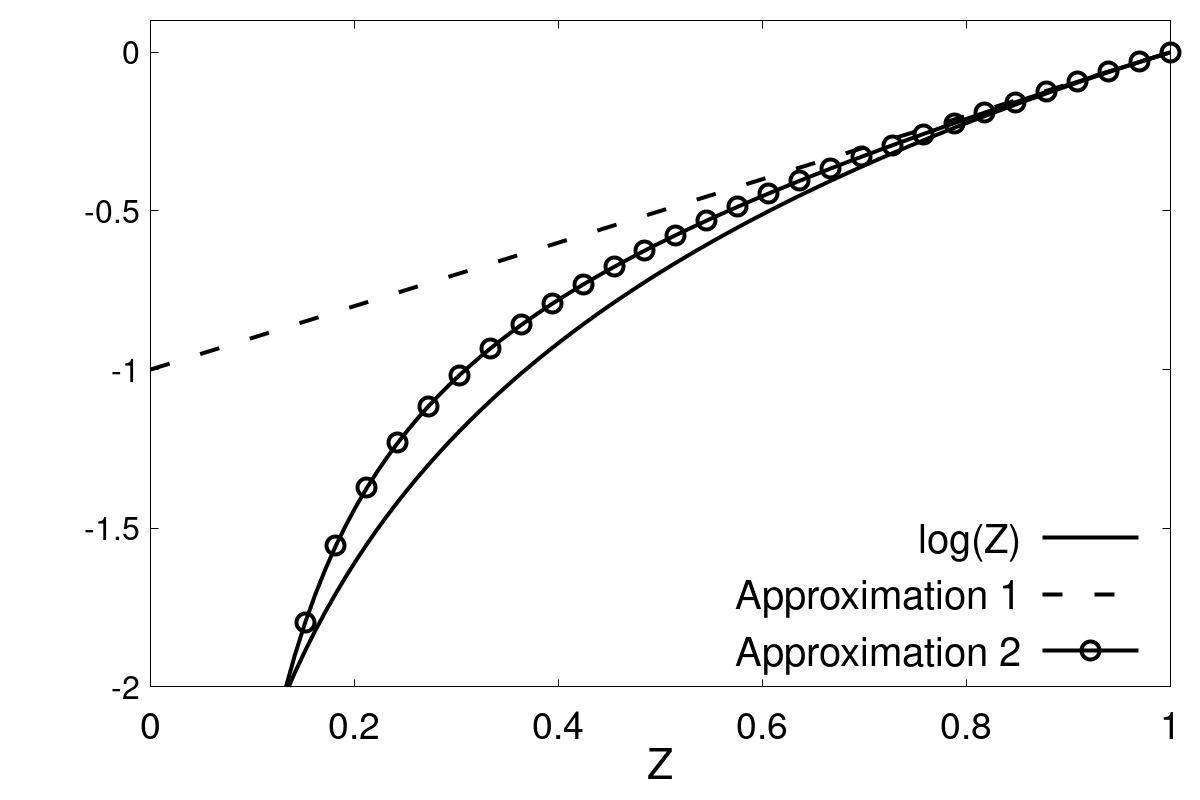}
	\caption{\label{fig1}Comparison of $\log(Z)$ with its two approximations. Approximation 1 being $\log(Z) \approx (Z-1)$ and approximation 2 being $\log(Z) \approx (Z-1)\left(w_1/Z + w_2 \right) $ with, $w_1=1/5$ and $w_2=4/5$.}
\end{figure}

The first order differential equations governing the dynamics of the system before lock down are:
\begin{align}
\label{Basic}
\begin{split}
\frac{d A}{dt} &= S  \, \alpha_0 \left(  I+   A\right) - \delta  A-\gamma A
\\
\frac{d   I }{dt} &=\delta  A-\gamma I \\
\frac{d S}{dt} &= -\alpha_0S  \, \left(I+A\right)\\
\frac{d R}{dt} &= \gamma \, (I+A) 
\end{split}
\end{align}
We define a new variable $M$, such that $M=A+I$. The dynamics of $M$ is given by
\begin{align}
\label{ApI}
\frac{dM}{dt} &= \alpha_0 \left(S M \right) - \gamma M 
\\
\frac{dM}{dS} &= \frac{1}{r_0 S} -1  
\end{align}
where $r_0$ is the basic reproduction number given by $r_0 = \alpha_0/\gamma$.
\begin{align}
	M = 1 - S + \frac{1}{r_0} \log \left(\frac{S}{S_{0}}\right)
\end{align}
where $S_0$ is the fraction of people who are susceptible at time(t)=0, and is a number very close to 1.
Using this relation in the evolution equation of S, gives:
\begin{align}
\frac{dS}{dt} = - S \alpha_0 \left( 1 - S + \frac{1}{r_0} \log \left(\frac{S}{S_{0}}\right) \right)
\end{align}
At this point, in order to extract a integrable exact solution, an approximation for the logarithm in the RHS is required. The two ways of approximating logarithm are
\begin{align}
\begin{split}
\text{Approximation 1} &: \qquad \log(Z) \approx Z-1
\\
\text{Approximation 2} &: \qquad \log(Z) \approx (Z-1) \left( \frac{w_1}{Z} +  w_2\right)
\end{split}  
\end{align}   
where, $w_1,w_2$are weights such that $w_1 + w_2 = 1$. 
\\
As $(S/S_{0})$ lies between $(0,1)$, we are only interested in $Z$ in the range $(0,1)$. The comparison between the two approximations is illustrated in FIG.\ref{fig1}.
Approximation 2, simplifies the differential equation to:
\begin{align}
\frac{dS}{dt} = \alpha_0 S^2  - \alpha_0 S - \frac{\alpha_0}{r_0} \left[\frac{S-S_0}{S_0}\right]\left(w_1 S_0 + w_2 S\right)
\end{align}
which can be simply written in the form
\begin{align}
\frac{dS}{dt} = a S^2 + bS + d
\end{align}
where, $ a = (S_0 r_0 -  w_2)/ r_0, \; 
b=   \left(r_0 +  w_1-  w_2\right)/r_0, \,  d=  w_1/{r_0}$.
\\
The solution upon integrating is
\begin{align}
\frac{1}{\sqrt{-b^2 + 4ad}} \left( 2 \tan^{-1} \left( \frac{b + 2as}{\sqrt{-b^2 + 4ad}} \right) \right) \Bigg |_{S_{0}}^{S} = t
\end{align}
where the integration variable is `$s$'.
As $b^2>4ad$, the equation can be rewritten as
\begin{align}
\frac{2}{\sqrt{b^2 - 4ad}}(-i) \> \> \arctan \left((i)\frac{-b-2as}{\sqrt{b^2-4ad}} \right) \Bigg |_{S_{0}}^{S} = t 
\end{align}
Using the identity $-i \arctan(ix) = \arctanh (x)$ 
\begin{align}
\frac{2}{\sqrt{b^2-4ad}} \>\> \arctanh \left(\frac{-b-2as}{\sqrt{b^2-4ad}} \right)\Bigg |_{S_{0}}^{S} = t
\end{align}
Using the identity: $\arctanh(x) = \frac{1}{2} \log \left(\frac{x+1}{x-1}\right)$
\begin{align}
\log \left( \frac{-b - 2as + \sqrt{b^2-4ad}}{-b-2as - \sqrt{b^2-4ad}} \right) \Bigg |_{S_{0}}^{S}  = \left( \sqrt{b^2-4ad}\right)t
\end{align}
Thus, using approximation 2 gives us an analytically tractable solution for the susceptible population
\begin{equation}
\frac{S}{S_0} = \frac{h   (1+ h_2 \exp(h \, \alpha_0 t)) }{2a   \left(1-h_2 \exp(h \, \alpha_0 t)\right)}
+\frac{b}{2a}
\end{equation}
where, $h_2 =    (2\, a-b-h)/(2\, a-b+h)$,  and $h,b$ are constants such that $h=\sqrt{b^2-4ad}$. 
  
\begin{figure}[h]
	\subfigure[Numerical solution of the SAIR system]{
	\includegraphics[scale=0.15]{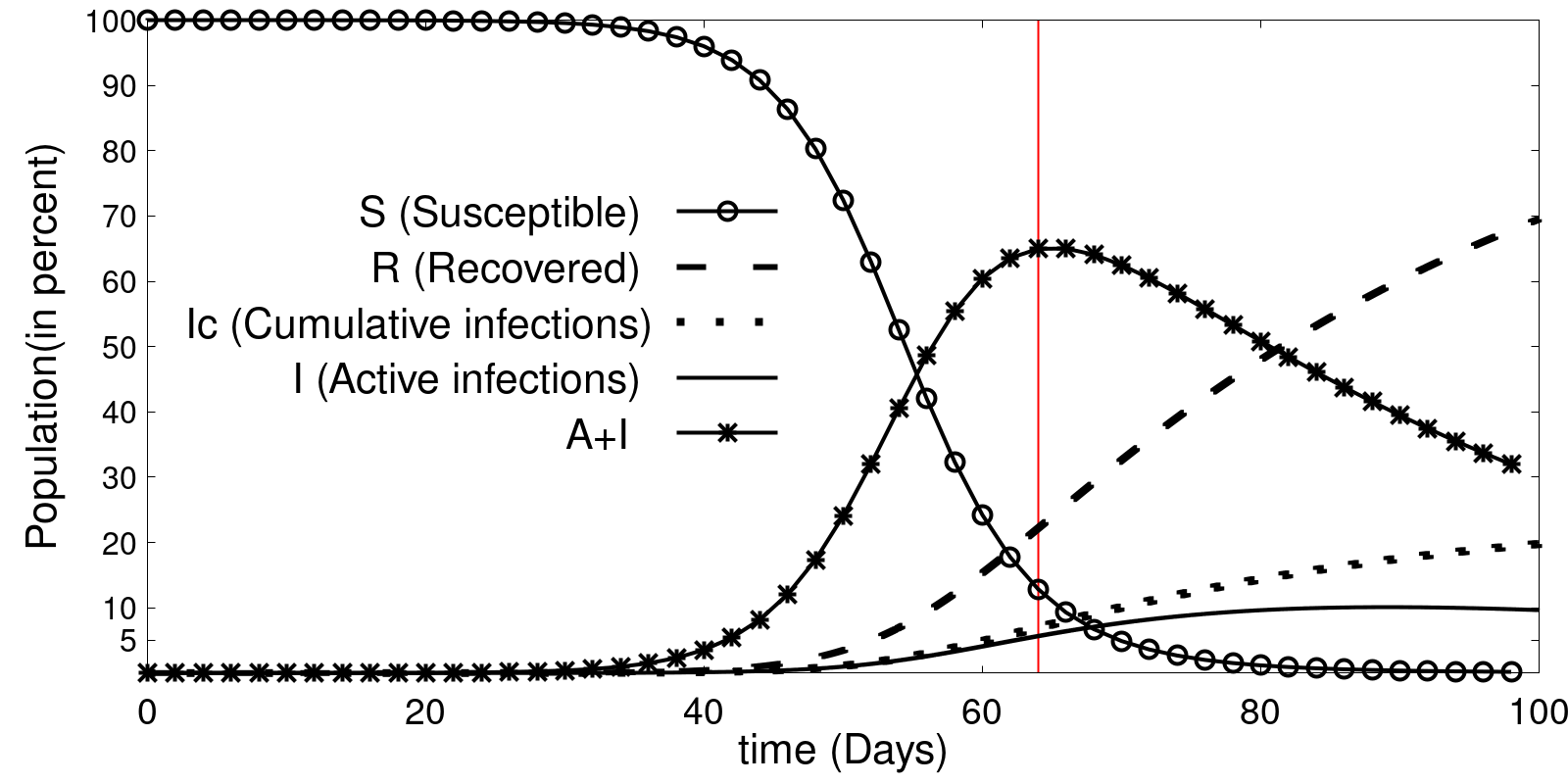}}
\subfigure[Analytical solution of the SAIR system using approximation 2]{
	\includegraphics[scale=0.15]{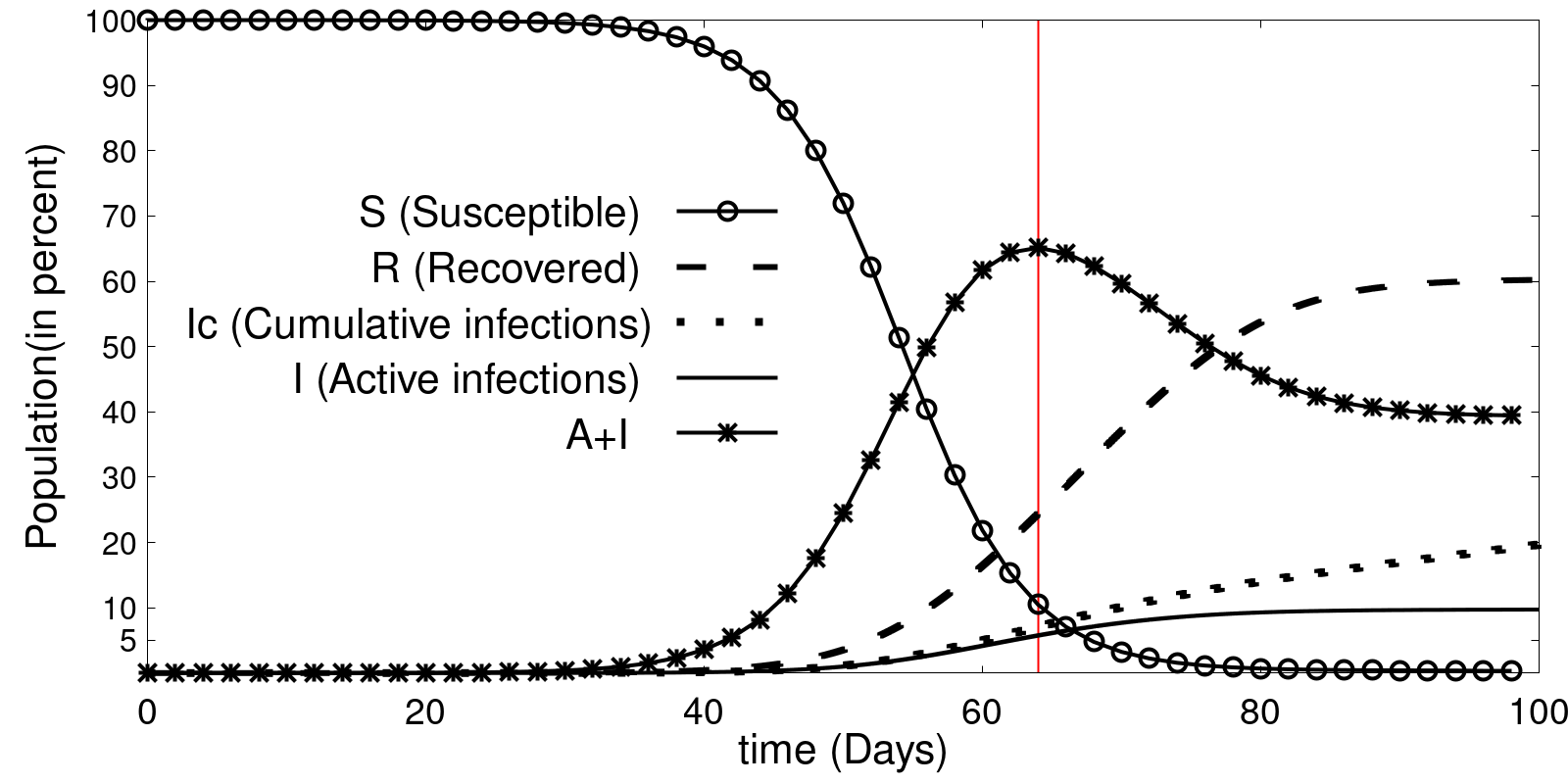}}
	\caption{\label{fig2} The figure illustrates the ability of the analytical solution found in section I, to correctly capture the dynamics of the SAIR model. The parameters used for these plots are, $\alpha=0.25, \gamma = 0.027, \delta_1 = 0.036$, which are in reasonable range of real time parameter values for COVID19 (discussed in section II). The initial conditions is one infected person in a million people. } 
\end{figure}

\section{II. Parameter Estimation}
\begin{figure}[h]
	\subfigure[]{
	\includegraphics[scale=0.16]{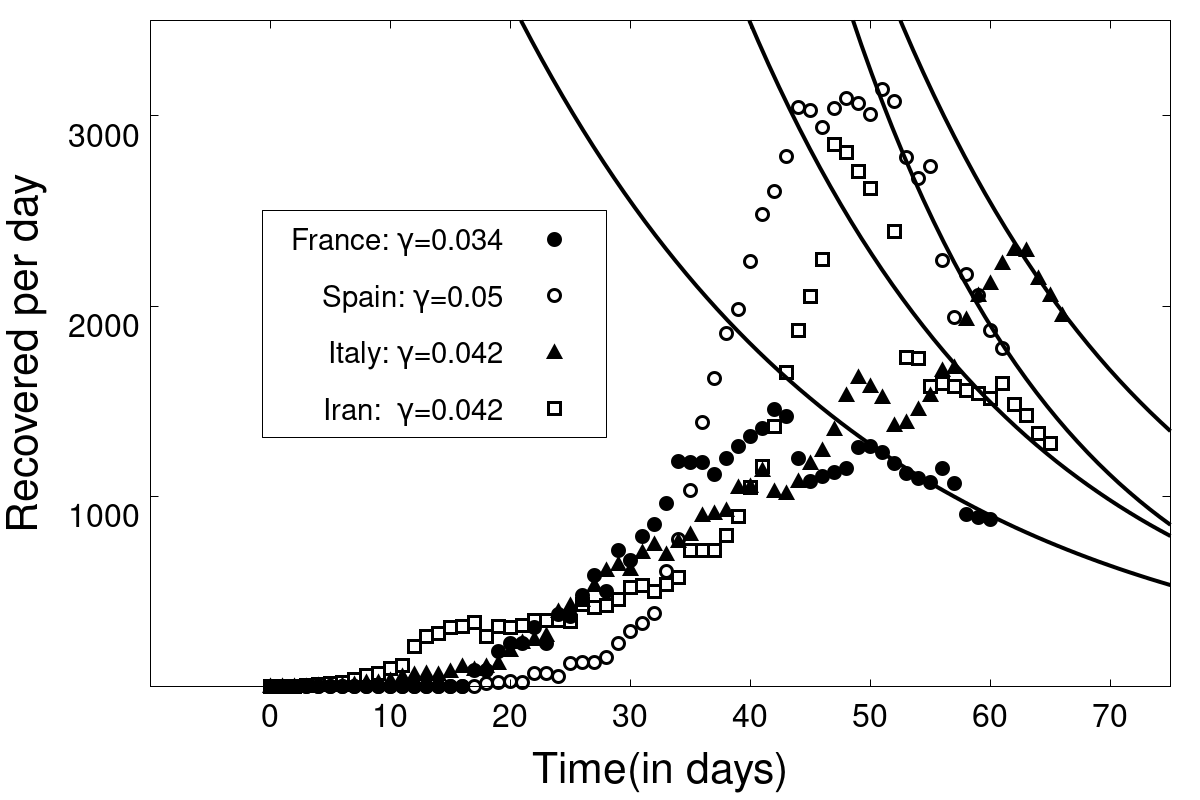}}
    \subfigure[]{
	\includegraphics[scale=0.16]{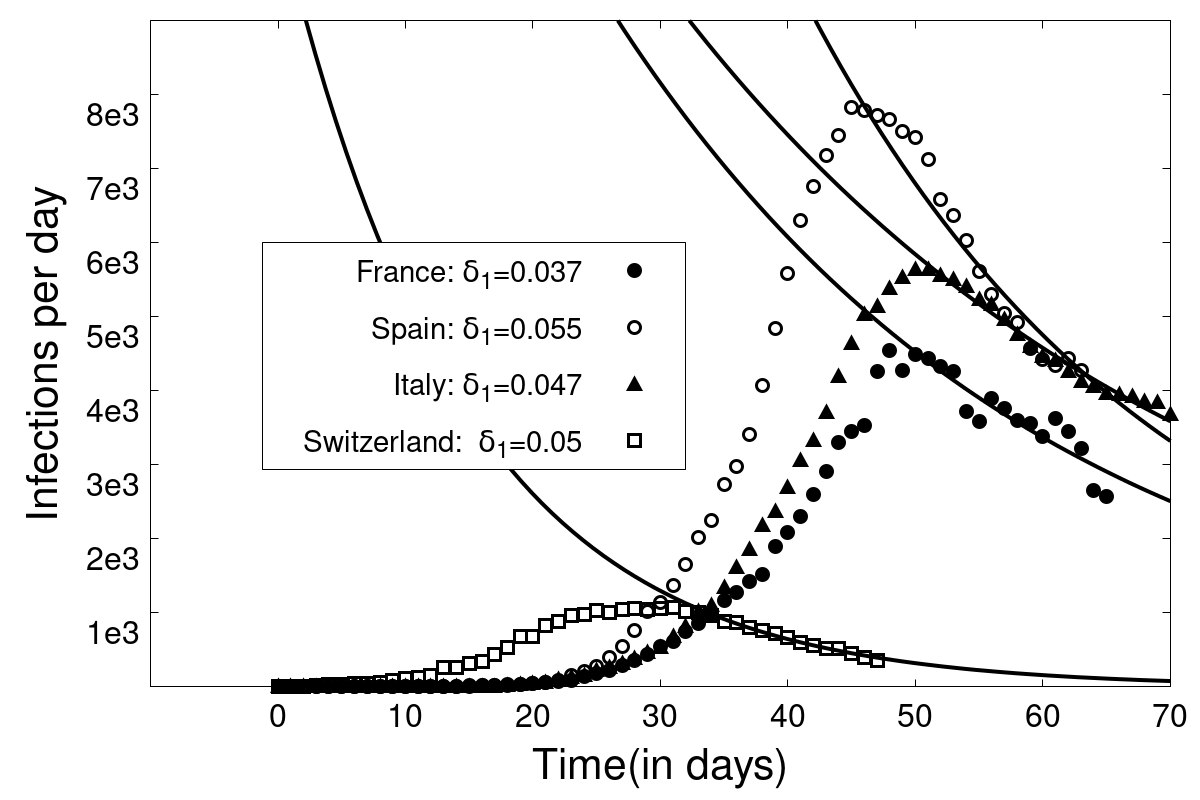}}
	\caption{\label{fig3}Estimation of parameters $\gamma$ and $\delta_1$ from post lock-down data}
\end{figure}

In this section we discuss the estimation procedure for the parameters ($\alpha_0, \gamma, \delta$). The analytical solutions for the infected(active) and recovered populations is known for both pre(discussed in section I) and post lock-down scenarios. These analytical solutions are then fit onto the real time data for several countries, to give us an estimate of the parameters relevant to COVID19. We begin with the post lock-down scenarios as the solutions are rather straight forward. After a '$\epsilon$' number of days post lockdown, the recovery rate is given by
\begin{equation}
\label{rdot_post}
\dot{R} =  \left[ \gamma (A+I)_{\rm lock + \epsilon} \right] \exp \{ -\gamma t\}
\end{equation}
and the infection is given by
\begin{equation}
\label{I_post}
I = \exp\{ -\delta_1 t \} \left[ I_{\rm lock+\epsilon} + \left( \frac{\delta(A+I)_{\rm lock+\epsilon}}{\delta_1-\gamma} \right) \left( \exp\{(\delta_1-\gamma)t \}-1\right)\right]
\end{equation}
Using Eq. \eqref{rdot_post} and real time recovery rate data for COVID19, the parameter $\gamma$ can be estimated as shown in FIG.\ref{fig3}a. Using Eq. \eqref{I_post} and real time infection rate data, parameter $\delta_1$ can be estimated as shown in FIG. \ref{fig3}b.   
\begin{figure}[h]
	\includegraphics[scale=0.16]{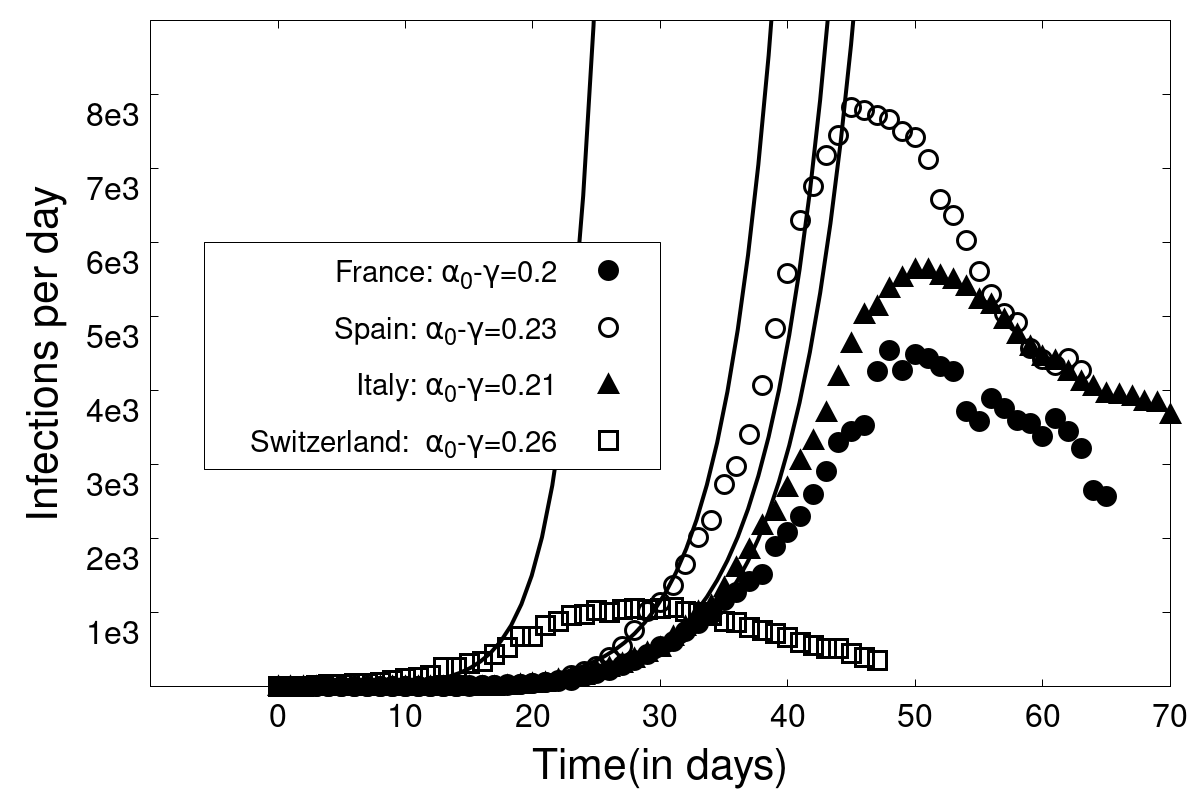} 
	\includegraphics[scale=0.16]{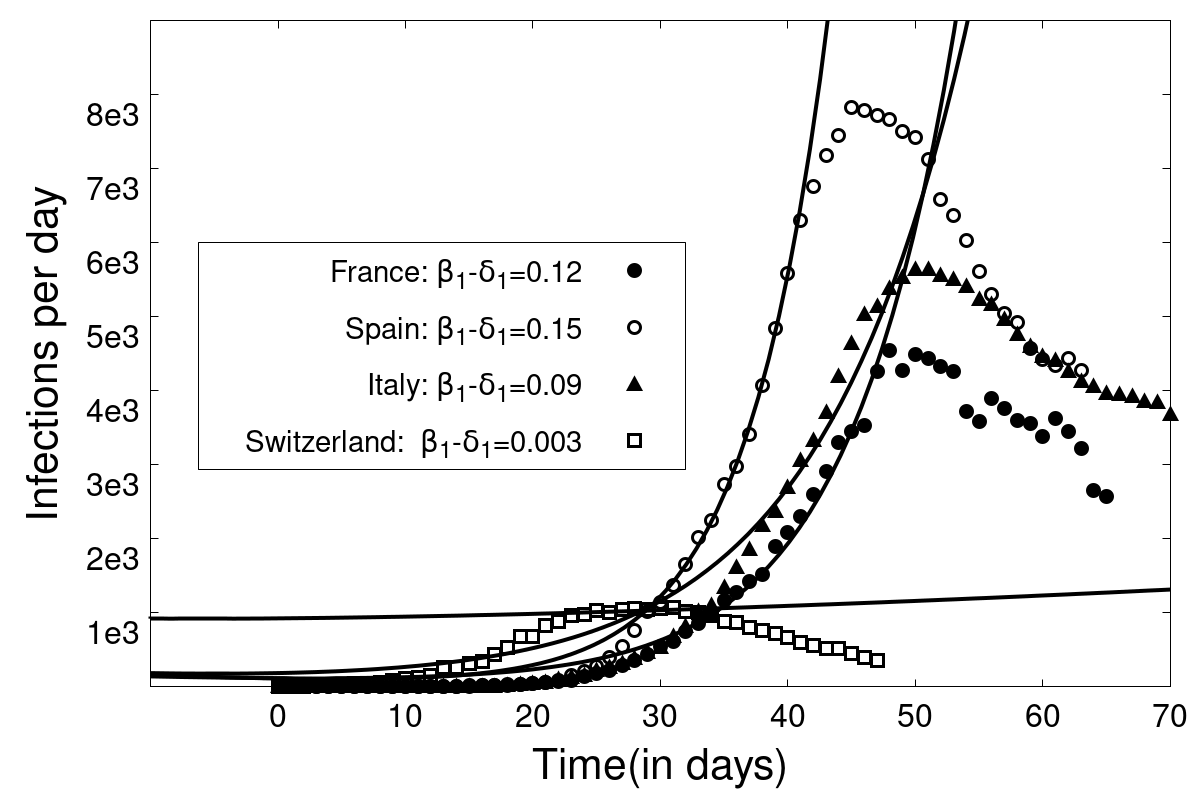}
	\caption{\label{fig4}Estimation of parameters $\alpha_0$ and $\beta_1$}
\end{figure}
The evolution of infections pre-lockdown and in early time limit is given by
\begin{align}
\label{pre_lock_infection}
I = \exp \{ -\delta_1 t \} \left[I_0 +  \int_0^t ds  \left( \frac{ \delta \exp\{ \delta_1 s \}}{ r_0} \right)
\left( k - g\tilde{S}(s) \right) \right]
\end{align}
where $k=(r_0-1)$ and $g=(S_0r_0-1)$. The solution post lock-down is given by
\begin{align}
\label{infections_post_lock}
I = \exp \{-\gamma t\}&(I_{\rm lock} - L) +  
\\
& L \exp\{ (-\delta_1 + \beta_1(1-{H}(t-\epsilon)))t \}
\end{align}
where, 
\begin{align}
L =  \frac{\delta \left( A_{\rm lock} \> \exp \{ \epsilon(\beta_1 - \delta_1) {H}(t-\epsilon) \} \right) } {(\gamma-\delta_1)+\beta_1(1-{H}(t-\epsilon))}
\end{align}
\begin{table} 
	\caption{\label{table1} Parameters extracted by fitting the solutions to the model we developed to the 3-day average data from the different countries.}
	\begin{ruledtabular}
		\begin{tabular}{|l|l|l|l|c|}
			Country & \pmb{$\alpha_0$} & \pmb{$\gamma$} & \pmb{$\delta_1$} &\pmb{$\beta_1$} 
			\\ \hline
			\textbf{France}    & $0.234 \pm 0.01$     & $0.034 \pm 0.002$ & $0.037 \pm 0.004$  & $0.15 \pm 0.007$         
			\\ 
			\textbf{Spain}     & $0.28 \pm 0.008$     & $0.05 \pm 0.003$   & $0.055 \pm 0.002$ & $0.2 \pm 0.01$                                    
			\\ 
			\textbf{Italy}     & $0.25 \pm 0.009$      & $0.042 \pm 0.002$  & $0.047 \pm 0.002$   &  $0.14 \pm 0.009$                                    
			\\ 
			\textbf{Switzerland} & $0.29 \pm 0.01$   &   $0.03 \pm 0.004$         & $0.05 \pm 0.003$     &  $0.053 \pm 0.004$                                  
			\\ 
		\end{tabular}
	\end{ruledtabular}
\end{table}
Now, using the above mentioned equations for infection rate and real time data for different countries, we estimate the parameters $\alpha_0$ and $\beta_1$, as shown in FIG.\ref{fig4}.
\begin{figure}
	\includegraphics[scale=0.15]{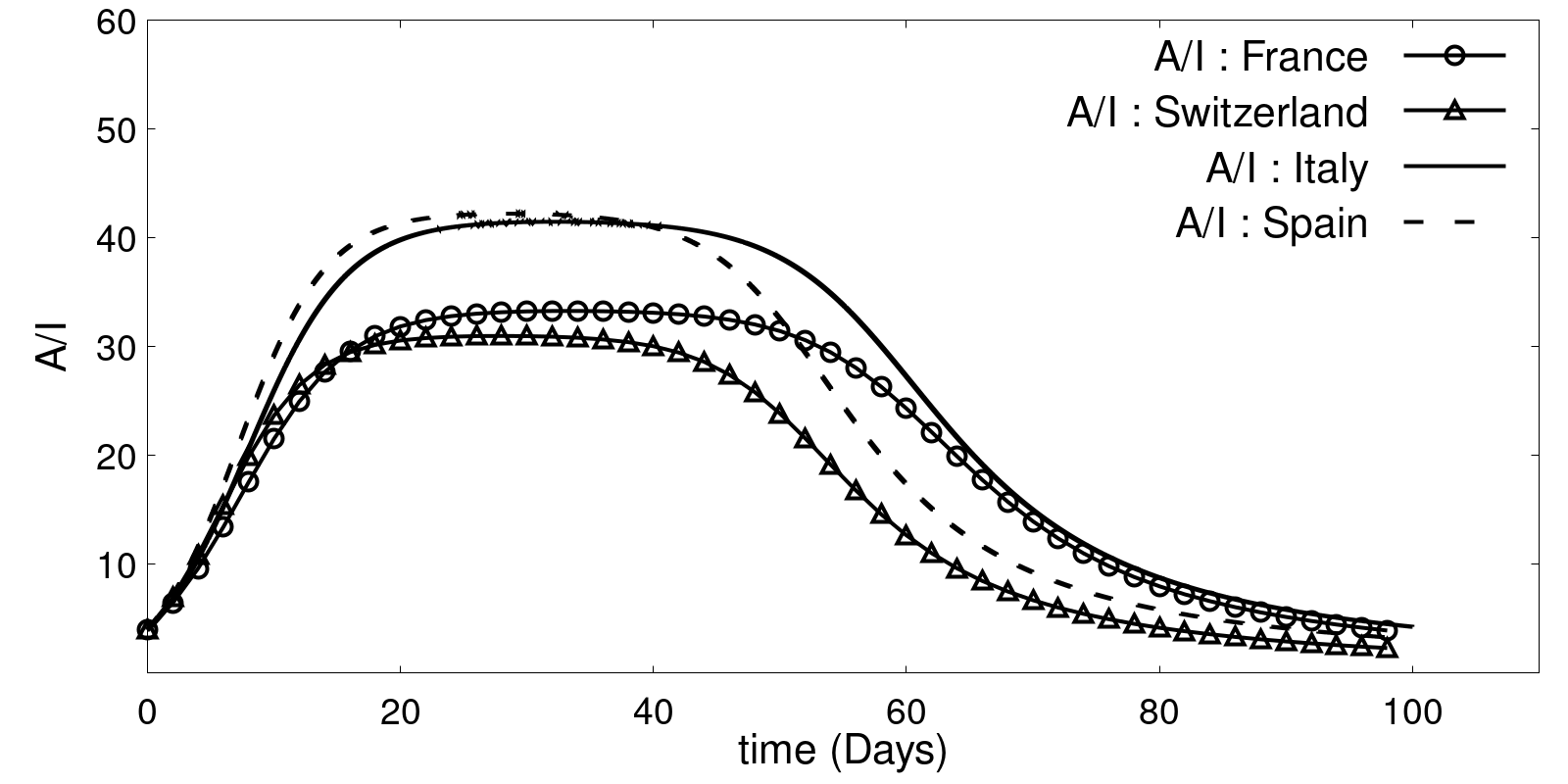}
	\caption{\label{fig5} The ratio of asymptomatics to the infected population as a function of time, and for a no-lockdown scenario. }
\end{figure}
\begin{table} 
	\caption{\label{table2} The details of the peak of infections extracted using relevant parameters for the COVID-19 dynamics in the different countries and under a hypothetical no-lockdown scenario}
	\begin{ruledtabular}
		\begin{tabular}{|l|l|l|l|c|}
			Country &  $I_{\rm max}$ & $(A+I)_{\rm max}$ & Ic when $I=I_{\rm max}$ & Ic when $(A+I)=(A+I)_{\rm max}$
			\\ \hline
			\textbf{France}    &  $6 \%$   & $56 \%$ & $12 \%$   &  $6 \%$       
			\\ 
			\textbf{Spain}     & $4.3 \%$     & $53 \%$   & $8.2 \%$  & $4.1 \%$
			\\
			\textbf{Italy} & $4 \%$  &   $51 \%$  &  $7.7 \%$    & $4 \%$  
			\\  
			\textbf{Switzerland} & $5.6 \%$   &    $52 \%$  &  $10 \%$    &  $4.5 \%$  
			\\ 
		\end{tabular}
	\end{ruledtabular}
\end{table}